\documentclass[journal,onecolumn]{IEEEtran}

\usepackage{cite}
\usepackage{amsmath}
\usepackage{amssymb}
\usepackage{graphics}
\usepackage{graphicx}
\usepackage{hyperref}
\usepackage{bm}
\usepackage{booktabs}
\usepackage{subcaption}

\newcommand\D{\ensuremath{\mathbf{D}} }
\newcommand\K{\ensuremath{\mathbf{K}} }

\newcommand\lp{\ensuremath{\left(}}
\newcommand\rp{\ensuremath{\right)}}

\newcommand\Reals{\ensuremath{\mathbb{R}} }
\newcommand\Complex{\ensuremath{\mathbb{C}} }

\newcommand\GP[1]{\ensuremath{ {\lp #1 \rp }} }

\newcommand\GB[1]{\ensuremath{ {\left[ #1 \right] }} }
\newcommand\GN[1]{\ensuremath{ {\left|\left| #1 \right|\right| }} }
\newcommand\GM[1]{\ensuremath{ {\left| #1 \right| }} }

\newcommand\GC[1]{\ensuremath{ {\left\{ #1 \right\} }} }
\newcommand\argmin{\ensuremath{\mathop{\text{argmin }} }}
\renewcommand\inf{\ensuremath{\mathop{\text{inf }} }}
\renewcommand\sup{\ensuremath{\mathop{\text{sup }} }}
\renewcommand\min{\ensuremath{\mathop{\text{min}} }}

\newcommand\argmax{\ensuremath{\mathop{\text{argmax }} }}

\newcommand\grad{\ensuremath{\nabla}}
\newcommand\hess{\ensuremath{\nabla^2}}
\newcommand\diag{\ensuremath{\mathop{\text{diag}} }}

\renewcommand\vec{\ensuremath{\mathop{\text{vec}} }}

\newcommand\elabel[1]{\label{eqn,#1}}
\newcommand\eref[1]{(\ref{eqn,#1})}

\newcommand\flabel[1]{\label{fig,#1}}
\newcommand\fref[1]{\ref{fig,#1}}

\newcommand\eg{{e.g.}, }

\newcommand\vs{{vs.} }

\newcommand\ie{{i.e.}, }

\newcommand\R{\ensuremath{\mathsf{R}} }

\newcommand\A{\ensuremath{{\mathbf{A}} }}

\newcommand\C{\ensuremath{{\mathbf{C}} }}
\newcommand\x{\ensuremath{{\mathbf{z}} }}

\newcommand\y{\ensuremath{{\mathbf{y}} }}

\newcommand\W{\ensuremath{\mathbf{W}} }
\newcommand\F{\ensuremath{\mathbf{F}} }

\newcommand\tr{\ensuremath{^\mathsf{T}} }
\renewcommand\th{\ensuremath{^\mathsf{H}} }

\newcommand\M{\ensuremath{{\mathbf{M}} }}

\newcommand\g{\ensuremath{{\mathbf{g}} }}

\newcommand\Ga{\ensuremath{{\mathbf{\Gamma}} }}

\renewcommand\H{\ensuremath{\mathbf{H}} }
\newcommand\U{\ensuremath{\mathbf{U}} }

\newcommand\ones{\ensuremath{{\bm{1}} }}
\newcommand\zeros{\ensuremath{{\bm{0}} }}

\newcommand\I{\ensuremath{\mathbf{I}} }

\renewcommand\u{\ensuremath{{\mathbf{u}} }}

\renewcommand\v{\ensuremath{{\mathbf{v}} }}
\renewcommand\d{\ensuremath{{\mathbf{d}} }}

\newcommand\eu{\ensuremath{{\bm{\eta_u}} }}

\newcommand\sn{\ensuremath{^{(n)}} }
\newcommand\snp{\ensuremath{^{(n+1)}} }

\newcommand\snz{\ensuremath{^{(0)}} }

\title{Algorithmic Design of Majorizers for Large-Scale Inverse Problems}
\author{Madison G. McGaffin \qquad \qquad Jeffrey A. Fessler \\
    Department of Electrical Engineering and Computer Science \\
    University of Michigan, Ann Arbor, MI
    \thanks{%
        This work is supported in part by NSF grant U01 EB 018753 and by
        equipment donations from Intel Corporation.
    } \\
    \vspace{1em}
    \today
}

\begin{document}

\maketitle{}

\begin{abstract}
Iterative majorize-minimize (MM) (also called optimization transfer) algorithms
solve challenging numerical optimization problems by solving a series of
``easier'' optimization problems that are constructed to guarantee monotonic
descent of the cost function.  Many MM algorithms replace a computationally
expensive Hessian matrix with another more computationally convenient
majorizing matrix.  These majorizing matrices are often generated using various
matrix inequalities, and consequently the set of available majorizers is
limited to structures for which these matrix inequalities can be efficiently
applied.  In this paper, we present a technique to algorithmically design
matrix majorizers with wide varieties of structures.  We use a novel
duality-based approach to avoid the high computational and memory costs of
standard semidefinite programming techniques.  We present some preliminary
results for 2D X-ray CT reconstruction that indicate these more exotic
regularizers may significantly accelerate MM algorithms.
\end{abstract}

\section{Introduction}

Given an $N \times N$ Hermitian matrix $\H$, we say that the Hermitian matrix
$\M$ majorizes $\H$ if none of the eigenvalues of $\M - \H$ are negative.
Matrix majorizers are central to majorize-minimize algorithms and are
ubiquitous in image processing
algorithms~\cite{chouzenoux:11:amm,muckley:15:fpm,erdogan:99:osa,ramani:13:ans}.
Better majorizers can significantly affect how quickly algorithms
converge~\cite{muckley:15:fpm,deleeuw:09:sqm,kim:13:aos}, but majorizers are
usually designed by hand on a per-problem basis. Algorithmic majorizer design
expands the class of usable majorizers and reveals more effective majorizers,
but a straightforward semidefinite programming approach requires too much
memory to be computationally feasible for many practical imaging problems.

This paper presents an algorithmic approach to designing a majorizing matrix
$\M$ for a general given Hermitian $\H$.  The algorithm we present has
relatively low memory requirements, and it is practical for large problems
where storing and manipulating dense $N \times N$ matrices would be infeasible.
The goal of this chapter is to enable the design of more exotic majorizers than
are currently accessible using various inequalities.  This expanded class of
majorizers may contain tighter majorizers~\cite{deleeuw:09:sqm} that lead to
faster convergence~\cite{fessler:93:ocd}.

Conventionally, a majorizing matrix $\M \succeq \H$ is found using a collection
of inequalities and matrix properties.  A simple and common bound is
$\M_\text{Lipschitz} = \lambda_\text{max}\GP{\H} \I$, which is often used in
optimization algorithms for which the cost function gradient is Lipschitz
continuous (with Lipschitz constant $\lambda_\text{max}\GP{\H}$).  This choice often is a
very loose bound.  Often a tighter bound is the diagonal matrix
\begin{align}
    \elabel{maj,eqn,sqs}
    \M_\text{SQS}
    &=
    \underset{j}\diag\GC{\sum_{i=1}^N \GM{\H}_{ij}}.
\end{align}
We call this the ``separable quadratic surrogates'' (SQS) majorizer due to its
ubituity in ordered subsets with SQS (OS-SQS) algorithms~\cite{erdogan:99:osa}.
If $\H$ contains only nonnegative entries (\eg $\A\tr\W\A$, the Hessian of the
X-ray CT data-fit term~\cite{thibault:07:atd,erdogan:99:osa}) then
$\M_\text{SQS}$ can be quickly computed via
\begin{align}
    \elabel{maj,eqn,sqs,compute}
    \M_\text{SQS}
    &=
    \underset{j}\diag\GC{\GB{\H \ones}_j}.
\end{align}
Our experiments (not shown) suggest that $\M_\text{SQS}$ is a fairly tight
diagonal majorizer when $\H$ contains only nonnegative entries, and its ease of
computation~\eref{maj,eqn,sqs,compute} makes it a very useful tool.  However,
when $\H$ contains negative entries, $\M_\text{SQS}$ appears to be less tight.
In this case, carefully designed majorizers that exploit the structure of $\H$,
\eg~\cite{muckley:15:fpm}, can significantly improve on $\M_\text{SQS}$.

This paper focuses on designing majorizers of the following form:
\begin{align}
    \elabel{maj,eqn,structure}
    \M = \K\th \D \K,
\end{align}
where $\D \in \Reals^{K \times K}$ is a diagonal matrix and $\K \in \Complex^{K
\times N}$, $K \ge N$ has full column rank.  We assume the matrix $\K$ is selected
by the MM algorithm designer
beforehand and focus on designing $\D$.  A special case is $\K = \I$,
which leads to a diagonal majorizer.
Our goal is to select the diagonal matrix $\D$ such that $\M$ is a majorizer
of $\H$ and such that the values of $\D$ are ``as small as possible'' so
that the majorizer is ``tight'' or ``sharp.''  To quantify this goal, we
pose the following convex optimization problem:
\begin{align}
    \M
    &=
    \K\tr \widehat \D \K, \qquad\text{where} \\
    \elabel{maj,eqn,cost}
    \widehat \D
    &=
    \argmin_{\D: \K\tr\D\K \succeq \H}
    \frac{1}{2}\GN{\d}_\W^2,
\end{align}
with positive diagonal weighting matrix $\W$.  The vector $\d \in \Reals^N$ is
the diagonal $\D$.

\section{Methods}

Let $\H \in \Complex^{N \times N}$ be a given Hermitian positive semidefinite
matrix, and let $\K \in \Complex^{K \times N}$ be a given matrix with full
column rank.  We aim to
majorize $\H$ with the matrix $\K\th \widehat \D \K$, where $\widehat \D$
solves
\begin{align}
    \elabel{problem}
    \widehat \D
    &=
    \argmin_{\D: \K\th\D\K \succeq \H} \frac{1}{2} \GN{\d}_\W^2.
\end{align}
The vector $\d \in \Reals^K$ is the diagonal of $\D$, and $\W \in \Reals^{K
\times K}$ is a given positive-definite weighting matrix.  This is a convex
minimization problem over a convex set, although the domain $\Omega$,
\begin{align}
    \Omega &= \GC{\d : \K\th\D\K \succeq \H}
\end{align}
is challenging to efficiently characterize.  The majorizer design problem we
pose~\eref{problem} could be solved using a semidefinite programming (SDP)
technique.  However, to the best of our knowledge, algorithms to
solve~\eref{problem} using SDP involve storing and manipulating dense $N \times
N$ matrices~\cite{boyd:04,fletcher:85:sdm}.  In practical image processing
problems, $N$ is the number of pixels in the image and can be very large, so
storing arbitrary $N \times N$ matrices is infeasible.

We rewrite the majorizer design problem~\eref{problem} using the
characteristic function $\iota_\Omega$ as
\begin{align}
    \elabel{problem,iota}
    \widehat \D
    &=
    \argmin_{\d \in \Reals^K} \frac{1}{2} \GN{\d}_\W^2 + \iota_\Omega\GP{\d}.
\end{align}
The characteristic function $\iota_\Omega\GP{\d}$ is zero when $\d \in \Omega$
(\ie{} $\K\th\D\K$ majorizes $\H$) and infinite otherwise.  Instead of directly
approaching~\eref{problem,iota}, we rewrite the characteristic function
$\iota_\Omega$ using a generalized convex conjugate~\cite{oettli:98:cff}:
\begin{align}
    \iota_\Omega\GP{\d}
    &=
    \sup_{\x \in \Complex^N} \x\th\GP{\H - \K\th\D\K}\x \\
    &=
    \sup_{\x \in \Complex^N} \x\th\H\x - \GP{\K\x}\th\D\GP{\K\x} \\
    &=
    \elabel{iota,dual}
    \sup_{\x \in \Complex^N} \x\th\H\x - \d\tr \GM{\K\x}^2,
\end{align}
where $\GM{\K\x}^2$ contains the elementwise squared complex moduli of $\K\x$:
\begin{align}
    \GM{\K\x}^2
    &=
    \underset{k}{\vec}\GC{\GM{\GB{\K\x}_k}^2}.
\end{align}
Substituting~\eref{iota,dual} in~\eref{problem,iota} yields the following
min-max problem for designing $\D$:
\begin{align}
    \elabel{saddle}
    \widehat \D &=
    \argmin_{\d \in \Reals^K} \sup_{\x \in \Complex^N} \mathsf{S}\GP{\d, \x}, \\
    \mathsf{S}\GP{\d, \x}
    &=
    \frac{1}{2} \GN{\d}_\W^2 - \d\tr\GM{\K\x}^2 + \x\th\H\x.
\end{align}
Reversing the order of the ``argmin'' and ``sup'' in~\eref{saddle} and solving
for the minimizing $\d$ in terms of $\x$ yields the
dual problem
\begin{align}
    \elabel{dual}
    \widehat \x
    &=
    \argmax_{\x \in \Complex^N}
    \mathsf{L}\GP\x, \qquad\text{where} \\
    \mathsf{L}\GP{\x}
    &=
    -\frac{1}{2}\GN{\GM{\K\x}^2}_{\W^{-1}}^2 + \x\th\H\x.
\end{align}
We solve this dual problem to recover the the optimal primal variable $\widehat
\D$.  This requires the following two results:
\begin{itemize}
    \item Strong duality, \ie{}
        \begin{align}
            \inf_{\d \in \Reals^K} \sup_{\x \in \Complex^N} \mathsf{S}\GP{\d, \x}
            &=
            \sup_{\x \in \Complex^N} \inf_{\d \in \Reals^K} \mathsf{S}\GP{\d, \x}.
        \end{align}
        Appendix~\ref{sec,strong,duality} provides a proof.

    \item A way to recover the primal variable $\widehat \D$ from a dual
        solution $\widehat \x$.  Appendix~\ref{sec,recover} shows
        \begin{align}
            \widehat \d
            &=
            \W^{-1} \GM{\K\widehat\x}^2,
        \end{align}
        where $\widehat \x$ solves the dual problem.
\end{itemize}
We have found a dual problem for the majorizer design problem~\eref{problem}.
Unlike the original problem, the dual problem does not directly use the
difficult-to-characterize domain $\Omega$ and can be approached without memory-
and computationally-expensive SDP techniques.  The drawback of this
transformation is that the dual problem~\eref{dual} is not concave.
Consequently, the steepest ascent algorithm in Section~\ref{sec,solve} may
converge to a local optimum.  Nonetheless these suboptimal solutions may be
useful, and Section~\ref{sec,maj} discusses modifications of
these local optima to provide a matrix that majorizes $\H$.

\subsection{Dual steepest ascent}
\label{sec,solve}

We find a local maximizer of the dual function $\mathsf{L}$ using steepest
ascent.  Initialize $\widehat \x\snz \ne \zeros$.  We compute the search
$\g\sn$ direction using the gradient of $\mathsf{L}\GP{\x\sn}$:
\begin{align}
    \g\sn
    &=
    \grad \mathsf{L}\GP{\x\sn} \\
    &=
    2\GP{\H\x - \K\th \W^{-1} \GM{\K\x}^2 \odot \K\x},
\end{align}
where $\odot$ is element-wise multiplication.  Maximizing the dual function
$\mathsf{L}$ along this search direction involves solving the following line
search problem:
\begin{align}
    \alpha\sn
    &=
    \argmax_{\alpha \in \Reals}
    f\sn\GP{\alpha}, \qquad\text{where} \\
    f\sn\GP{\alpha}
    &=
    \mathsf{L}\GP{\x\sn + \alpha \g\sn}.
\end{align}
Setting $\GP{f\sn}'\GP{\alpha} = 0$ yields
\begin{align}
    \elabel{roots}
    0
    &=
    c_3 \alpha^3 + c_2 \alpha^2 + c_1 \alpha + c_0, \qquad \text{where} \\
    c_3 &= 2\v_2\tr\W^{-1}\v_2 \\
    c_2 &= 2\v_2\tr\W^{-1}\v_1 \\
    c_1 &= 2\v_2\tr\W^{-1}\v_0 + \v_1\tr\W^{-1}\v_1 - 2 b_2 \\
    c_0 &= \v_1\tr\W^{-1}\v_0 - b_1,
\end{align}
\begin{align}
    \v_2 &= \GM{\K\g\sn}^2
        & b_2 &= \GP{\g\sn}\th\H\g\sn \\
    \v_1 &= 2\cdot\text{Real}\GC{\K\g\sn \odot \overline{\K\x\sn}}
        & b_1 &= 2\cdot\text{Real}\GC{\GP{\g\sn}\th\H\x\sn} \\
    \v_0 &= \GM{\K\x\sn}^2.
\end{align}
The root-finding problem~\eref{roots} can be solved using an off-the-shelf
routine, \eg{} {\tt roots} in the software package {\tt octave}.  We loop
over the real roots of $\GP{f\sn}'\GP{\alpha}$ and choose the step length
$\alpha\sn$ that maximizes the dual function.

Each iteration of the proposed algorithm requires one multiplication by $\H$
and one multiplication by $\K$ and $\K\th$, and the algorithm stores only
length-$N$ and -$K$ vectors.  Thus the algorithm is practical even for large-scale
problems like CT image reconstruction where one must compute $\H\x$ on the
fly.  This procedure will find a local maximizer of the
dual function; the next section discusses how to manipulate the resulting
approximate solution to produce a majorizer of $\H$.

\subsection{Ensuring majorization}
\label{sec,maj}

The dual problem~\eref{dual} is not concave, so the steepest ascent algorithm
in this section will not in general converge to a global maximizer of the
dual function $\mathsf{L}$.  This means that the induced matrix,
\begin{align}
    \K\th\widetilde \D \K
    &=
    \K\th \W^{-1} \diag_{i}\GC{\GB{\GM{\K\widetilde\x}^2}_i} \K,
\end{align}
may not majorize $\H$.  To compensate for this possible suboptimality, we scale
the resulting matrix:
\begin{align}
    \widetilde \M
    &=
    \alpha \K\tr\widetilde\D\K \succeq \H,\qquad\text{where} \\
    \alpha
    &\in
    \GB{%
        \lambda_\text{max}\GP{%
            \GP{\K\th\widetilde\D\K}^{-\frac{1}{2}}
            \H
            \GP{\K\th\widetilde\D\K}^{-\frac{1}{2}}
        },
        3
    }.
\end{align}
This results in a majorizer for $\H$; see Appendix~\ref{sec,scaling} for a
proof.  If $\GP{\K\th\widetilde\D\K}^{-\frac{1}{2}}$ is easily computed, then
power iteration can be used to find the optimal (minimal) of $\alpha$.  This will be the case when \eg{}
$\K$ is a unitary matrix.  If $\K$ is more complicated, power iteration may be
undesirably computationally expensive, and we instead use the looser scaling
$\alpha = 3$.

\section{Preliminary experiments}
\renewcommand\x{\ensuremath{{\mathbf{x}} }}

\subsection{Weighted Toeplitz matrices and circulant majorizers}
\label{weighted,toeplitz}

\begin{figure}
    \centering
    \begin{subfigure}{.3\textwidth}
        \centering
        \includegraphics[width=\textwidth]
                {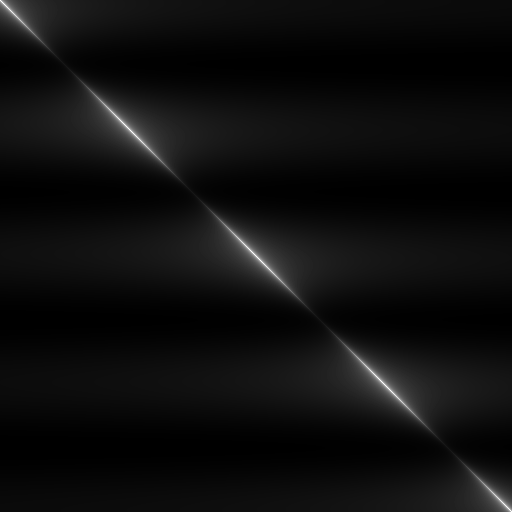}
        \caption{$\F$}
        \label{maj,fig,f}
    \end{subfigure}
    \begin{subfigure}{.3\textwidth}
        \centering
        \includegraphics[width=\textwidth]
                {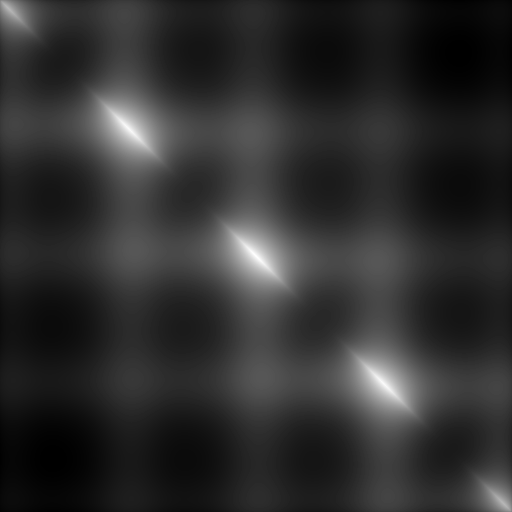}
        \caption{$\H$}
        \label{maj,fig,h}
    \end{subfigure}
    \begin{subfigure}{.3\textwidth}
        \centering
        \includegraphics[width=\textwidth]
                {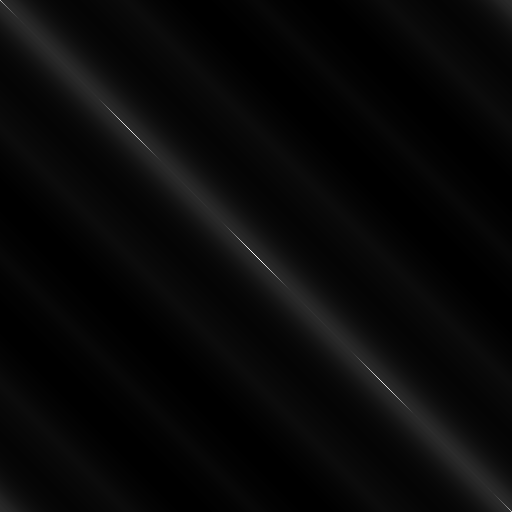}
        \caption{$\M_\text{Design-Circ+Diag}$}
        \label{maj,fig,cd}
    \end{subfigure}
    \caption{The nonnegative matrices $\F$ and $\H$ from the experiment in
        Section~\ref{weighted,toeplitz}, and the matrix
        $\M_\text{Design-Circ+Diag}$ produced by our proposed algorithm.}
    \label{maj,fig,small}
\end{figure}

We generated the $N \times N$ weighted Toeplitz matrix $\F$ with
$N = 512$ and entries
\begin{align}
    \GB{\F}_{ij}
    &=
    \frac{0.1 + \cos^2\GP{2\pi \frac{i}{N}} }{\sqrt{1 + \GM{i - j}} }
\end{align}
and set $\H = \F\tr\F$.  This choice is inspired by the $1/r$-like response
of the CT system matrix~\cite{clinthorne:93:pmf}.
Figures~\ref{maj,fig,f} and~\ref{maj,fig,h} show
$\F$ and $\H$, respectively.  We generated three diagonal majorizers:
\begin{itemize}
    \item $\M_\text{Lipschitz} = \lambda_\text{max}\GP{\H} \I$,
    \item $\M_\text{SQS}$ using~\eref{maj,eqn,sqs,compute}, and
    \item $\M_\text{Design-Diag}$, using the algorithing proposed in this
        chapter with $\K = \I$.
\end{itemize}
We also generated $\M_\text{Design-Circ+Diag}$, a combination of circulant
and diagonal matrices, using the proposed algorithm with
\begin{align}
    \K_\text{Circ+Diag}
    &=
    \GB{%
        \begin{matrix}
            \U_\text{DFT} \\
            \I
        \end{matrix}
    }.
\end{align}
Finally, we computed $\M_\text{Circ}$
\begin{align}
    \M_\text{Circ} &= \beta \widehat \C,
\end{align}
where $\widehat \C$ is the best circulant approximation to $\F$ in the Frobenius-norm
sense~\cite{chan:88:aoc},
\begin{align}
    \widehat \C &= \argmin_{\C~\text{circulant}} \GN{\C - \H}_\F^2,
\end{align}
and $\beta$ is chosen with power iteration so $\beta \widehat \C \succeq \H$.

\begin{figure}
    \centering
    \begin{subfigure}{.45\textwidth}
        \centering
        \includegraphics[width=\textwidth]
                {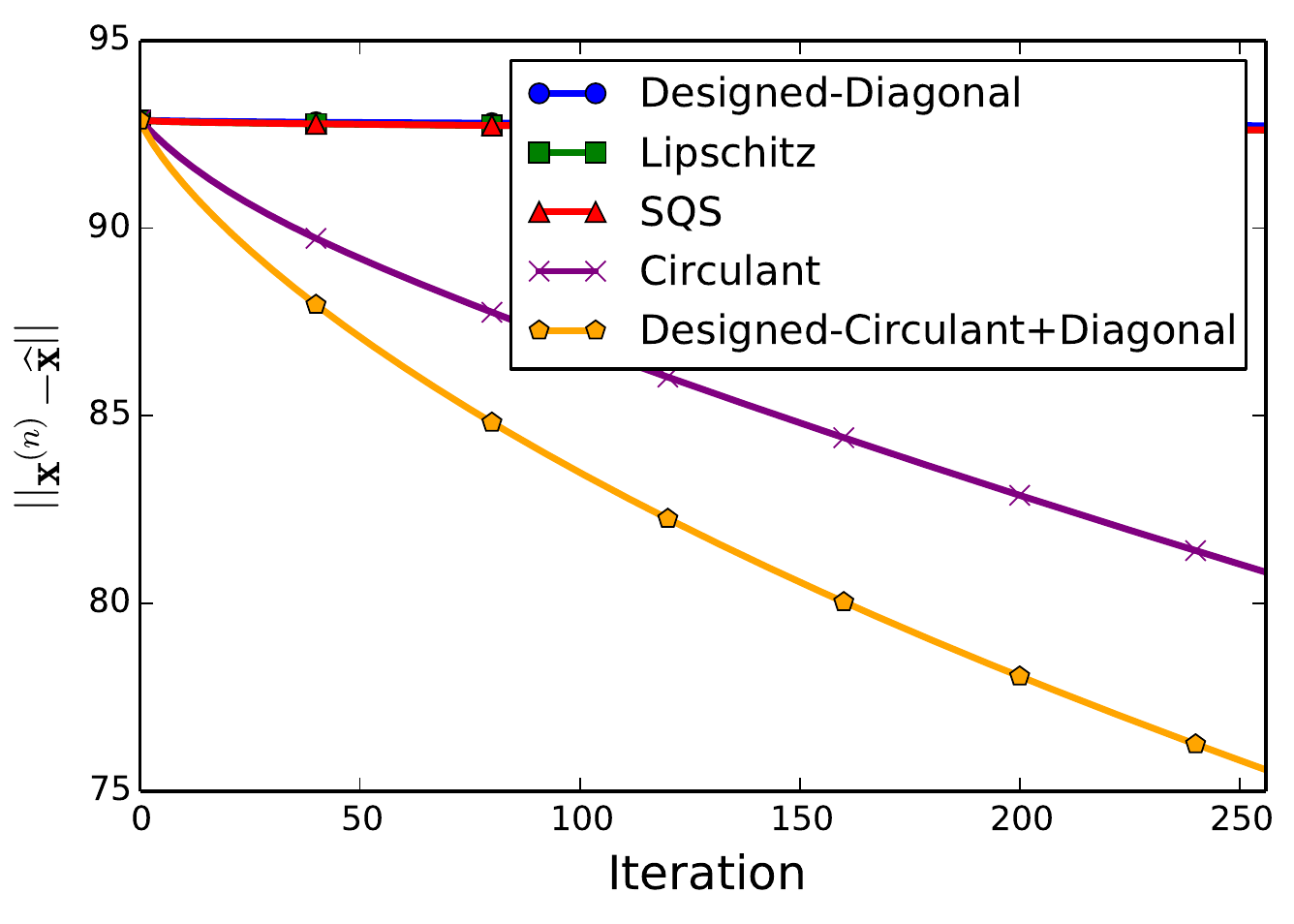}
        \caption{Distance to solution}
        \label{maj,fig,cost}
    \end{subfigure}
    \begin{subfigure}{.45\textwidth}
        \centering
        \includegraphics[width=\textwidth]
                {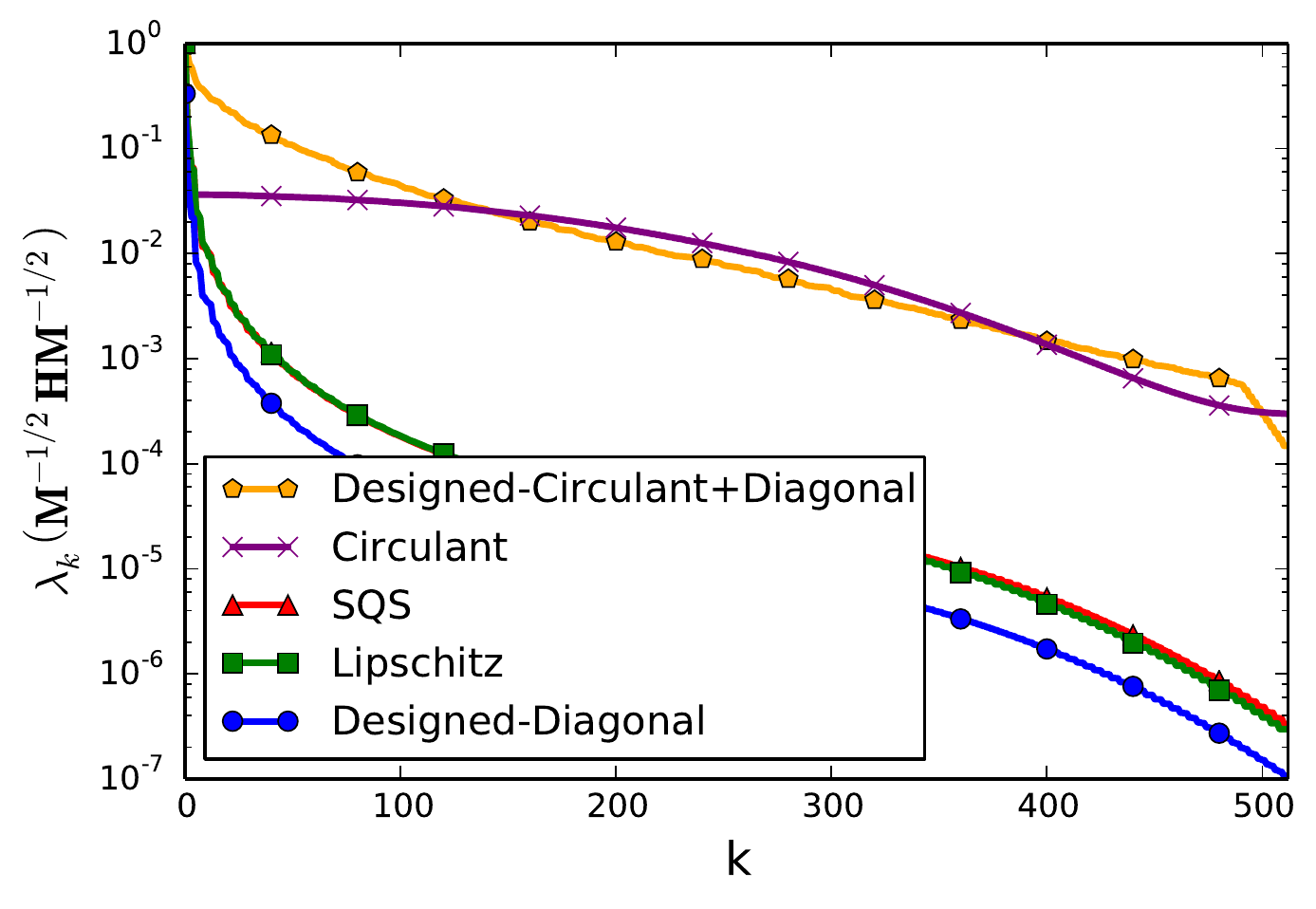}
        \caption{Majorized spectra}
        \label{maj,fig,spectra}
    \end{subfigure}
    \caption{Convergence plots and majorized spectra for the small Toeplitz
        experiment in Section~\ref{weighted,toeplitz}.  The partially circulant
        majorizer $\M_\text{Design-Circ+Diag}$ acts like both a majorizer and a
        preconditioner, accelerating convergence of the simple
        majorize-minimize algorithm.  The ideal majorizer inverts the
        matrix $\H$ and produces a uniform majorized spectrum with value 1.}
    \label{maj,fig,small,plots}
\end{figure}

We used each of the majorizers to solve the quadratic minimization
problem
\begin{align}
    \elabel{maj,eqn,small,mm}
    \widehat \x
    &=
    \argmin_\x \frac{1}{2} \x\tr\H\x + \x \tr \g,
\end{align}
with $\g$ and $\x\snz$ initialized with zero-mean normal random values.
We performed the following simple majorize-minimize (MM) procedure:
\begin{align}
    \elabel{maj,eqn,small,mm2}
    \x\snp
    &=
    \x\sn - \M^{-1}\GP{\H\x\sn + \g}.
\end{align}
This experiment explores the relative accelerations that different majorizers
provide.  To solve~\eref{maj,eqn,small,mm} even faster with a majorize-minimize
algorithm, we would also use some first-order acceleration
scheme~\cite{nesterov:83:amf,nesterov:05:smo,kim:15:ogm}.
Figure~\ref{maj,fig,cost} shows how quickly the MM algorithm iterates
$\GC{\x\sn}$ converged to the solution of~\eref{maj,eqn,small,mm} as a function
of iteration with each of the majorizers, and Figure~\ref{maj,fig,spectra}
shows the eigenvalues of $\M^{-\frac{1}{2}} \H \M^{-\frac{1}{2}}$ for each
majorizer $\M$.  For fast convergence, ideally those eigenvalues would be near
$1$~\cite{fessler:93:ocd}.

The designed diagonal majorizer underperforms the more conventional SQS
majorizer, $\M_\text{SQS}$.  This is possibly due to the proposed algorithm
producing a suboptimal solution to the majorizer design dual problem.
Regardless, diagonal majorizers are not our primary interest here: the
proposed algorithm is more useful for generating majorizers with more exotic
structures.

Except for edge conditions, circulant matrices and Toeplitz matrices are very
similar.  Because $\M_\text{Circ}$ and $\M_\text{Design-Circ+Diag}$ contain
circulant terms, they can approximate $\H$ while majorizing it.  When these
matrices are inverted in the majorize-minimize
procedure~\eref{maj,eqn,small,mm2}, they act as both preconditioners and
majorizers.  The preconditioning effect appears in better-conditioned
spectra (\ie{} closer to unform 1s) for $\M_\text{Circ}$ and
$\M_\text{Design-Circ+Diag}$ in Figure~\ref{maj,fig,spectra} and in faster
convergence rates in Figure~\ref{maj,fig,cost}.  Because the majorizer design
algorithm in this chapter can generate a majorizer with both circulant and
diagonal components, it can capture the nonuniform weighting in $\H$ better than
$\M_\text{circ}$.  This results in further improvement over $\M_\text{circ}$.

\subsection{X-ray CT reconstruction}

Consider the following unconstrained X-ray CT reconstruction problem:
\begin{align}
    \elabel{xray,recon}
    \widehat \x
    &=
    \argmin_\x
    \frac{1}{2}\GN{\A\x - \y}_\W^2 + \R\GP{\x}
\end{align}
with X-ray CT system matrix $\A$, noisy measurements $\y$, diagonal matrix of
positive statistical weights $\W$ and edge-preserving regularizer
$\R$~\cite{thibault:07:atd}.  In this experiment, we assume the edge-preserving
regularizer $\R$ is differentiable.  The statistical weights $\W$ contain
patient-specific data that one can separate from the large CT system matrix
with variable splitting:
\begin{align}
    \widehat \x
    &=
    \argmin_\x \frac{1}{2}\GN{\u - \y}_\W^2 + \R\GP\x \qquad\text{such that $\u = \A\x$.}
\end{align}
Applying the alternating directions method of multipliers (ADMM) to this
constrained problem yields the following set of iterated
updates~\cite{ramani:12:asb,nien:15:fxr} with positive-definite penalty matrix
$\Ga$ and dual variable $\eu$:
\begin{align}
    \elabel{admm,u}
    \u\snp
    &=
    \GP{\W + \Ga}^{-1}\GP{\W\y + \Ga\GP{\A\x\sn + \eu\sn}} \\
    \elabel{admm,mu}
    \x\snp
    &=
    \argmin_\x \frac{1}{2}\GN{\A\x - \u\snp + \eu\sn}_\Ga^2 + \R\GP\x \\
    \elabel{admm,eu}
    \eu\snp
    &=
    \eu\sn + \A\x\snp - \u\snp.
\end{align}
We can choose $\Ga$ to make the $\u$ update~\eref{admm,u} easy, and the dual
variable update~\eref{admm,eu} is trivial.  For this experiment, we follow
the guidance in~\cite{ramani:12:asb} and set
\begin{align}
    \Ga &= \underset{i}{\text{median}}\GC{w_i} \cdot \I.
\end{align}
The only challenging operation in this algorithm is the $\x$
update~\eref{admm,mu} involving the computationally expensive CT system matrix
$\A$.

Let $\M \succeq \A\tr\Ga\A$ be a majorizer for the quadratic term in Hessian of
the $\x$ update~\eref{admm,mu}.  Instead of solving~\eref{admm,mu} exactly,
we majorize the quadratic term and descend the surrogate function
\begin{align}
    \x\snp
    &\approx
    \argmin_\x \frac{1}{2}\GN{\x - \x\sn}_\M^2 + \GP{\x - \x\sn}\tr\A\tr\Ga\GP{\A\x\sn - \u\snp + \eu\sn} + \R\GP\x
\end{align}
using five iterations of conjugate gradients.

We simulated a noisy 2D fan-beam scan of an XCAT~\cite{segars:08:rcs} phantom
with an 888-channel detector and 984 views, and then reconstructed images onto
a $512 \times 512$-pixel grid.  Figure~\fref{ct,fbp} shows the intial image
$\x\snz$ from filtered backprojection.

\begin{figure}
    \centering
    \begin{subfigure}{.45\textwidth}
        \includegraphics[width=\textwidth]{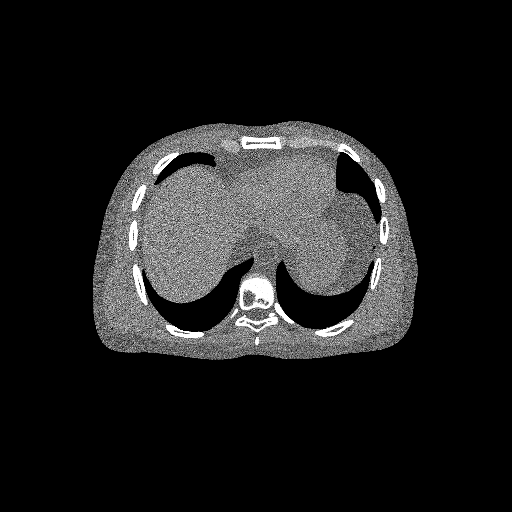}
        \caption{$\x\snz$}
        \flabel{ct,fbp}
    \end{subfigure}
    \begin{subfigure}{.45\textwidth}
        \includegraphics[width=\textwidth]{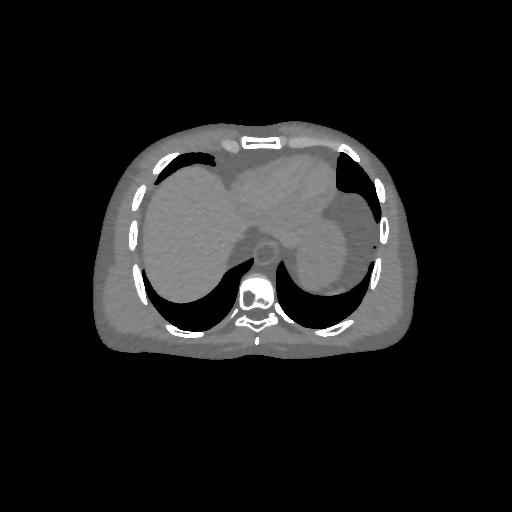}
        \caption{$\x^{(64)}_\text{Down}$}
        \flabel{ct,ref}
    \end{subfigure}
    \caption{Initial image from filtered backprojection~(\fref{ct,fbp}) and
        the output of the ADMM algorithm with majorizer $\M_\text{Down}$
        after 64 iterations~(\fref{ct,ref}).}
\end{figure}

\begin{figure}
    \centering
    \begin{subfigure}{.45\textwidth}
        \includegraphics[width=\textwidth]{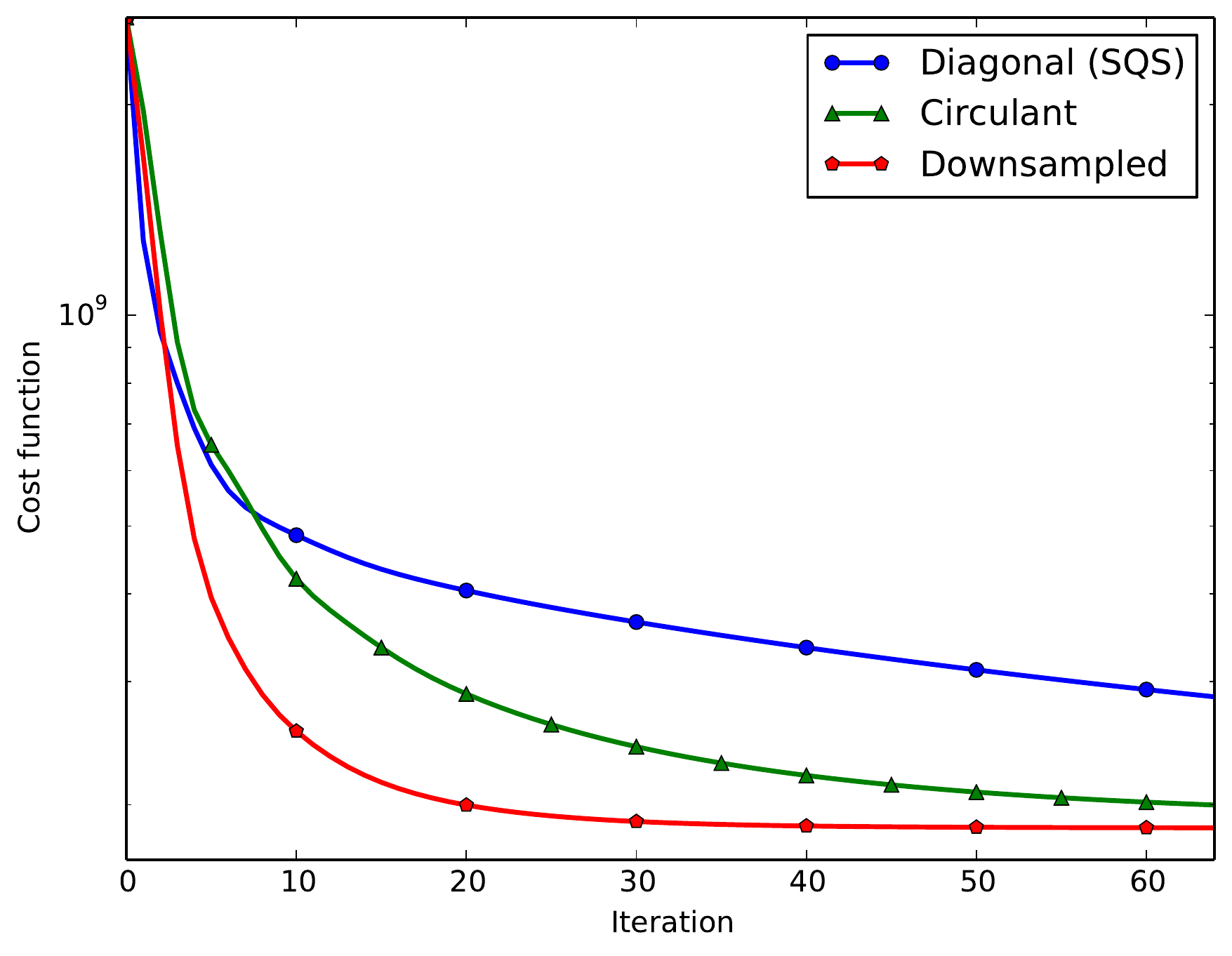}
        \caption{Cost function \vs iteration}
        \flabel{ct,iter}
    \end{subfigure}
    \begin{subfigure}{.45\textwidth}
        \includegraphics[width=\textwidth]{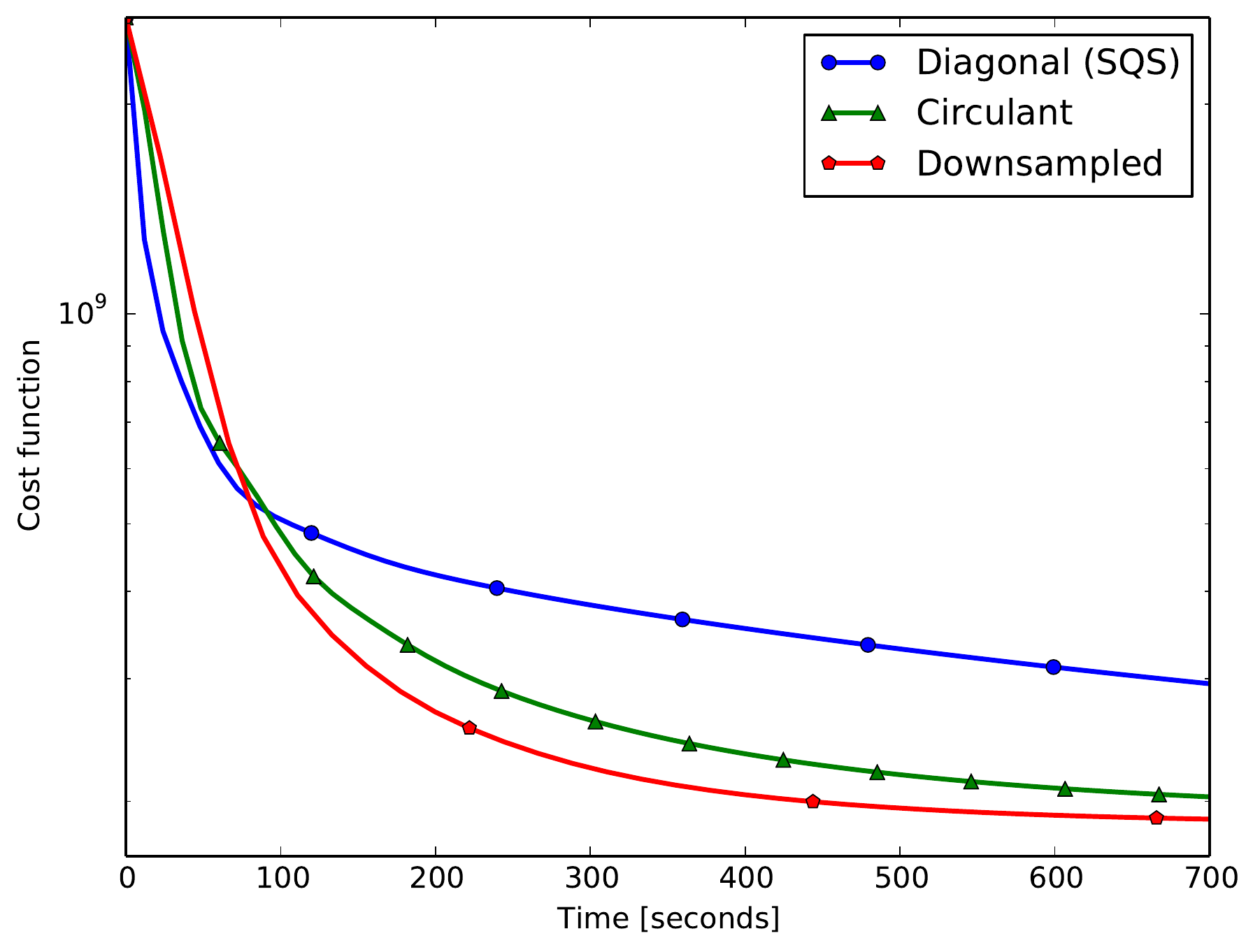}
        \caption{Cost function \vs time}
        \flabel{ct,time}
    \end{subfigure}
    \caption{Per-iteration~(\fref{ct,iter}) and per-time~(\fref{ct,time})
        cost function values for the ADMM algorithm using the three majorizers
        $\M_\text{SQS}$, $\M_\text{Circ}$ and $\M_\text{Down}$.}
\end{figure}

We ran the ADMM algorithm~\eref{admm,u}-\eref{admm,eu} with the following
majorizers $\M$ for the $\x$ update:
\begin{itemize}
    \item the diagonal majorizer $\M_\text{SQS}$~\eref{maj,eqn,sqs};
    \item a circulant majorizer $\M_\text{Circ}$ found by the proposed majorizer
        design algorithm with $\K = \U_\text{DFT}$; and
    \item a majorizer using a ``downsampled'' version of the CT Gram matrix
        $\A\tr\A$, $\M_\text{Down}$.  We set
        \begin{align}
        \K_\text{Down} &= \GB{\begin{matrix} \A_\text{Down} \\ \I \end{matrix}},
        \end{align}
        and used the proposed majorizer design algorithm, so $\M_\text{Down}$
        was a sum of a diagonal matrix and an approximation to the CT Gram
        matrix.  For this experiment, $\A_\text{Down}$ was a fan-beam CT system
        matrix that matched $\A$ except it had 82 (instead of 984) views and
        128 (instead of 888) channels.  The downsampled $\A_\text{Down}$ system had the same spatial
        coverage as the full system $\A$ but lower computational cost.
\end{itemize}
We ran the majorizer design algorithm for 128 iterations for both
$\M_\text{Circ}$ and $\M_\text{Down}$, and used the looser $\alpha = 3$ scaling
to ensure majorization (see Section~\ref{sec,maj}).

All algorithms were run on a machine with a 12-core Intel Xeon processor, and
multiplications with the system matrix $\A$ were performed using multithreaded
C code.  Figure~\fref{ct,ref} shows the output of ADMM algorithm using
$\M_\text{Down}$ after 64 iterations.  This is a preliminary experiment, and we
do not yet have an essentially converged reference image to which to compare
the tested algorithms, so we use the cost function at each iteration instead.
Figures~\fref{ct,iter} and~\fref{ct,time} show how quickly each algorithm
descends the cost function~\eref{xray,recon} as functions of iteration and
time, respectively.

Because the designed majorizers $\M_\text{Circ}$ and $\M_\text{Down}$ capture
more of the structure of the Hessian $\H = \A\tr\A$, they serve as both
majorizers and preconditioners in the $\x$ update~\eref{admm,mu}.  This
allows the ADMM algorithm using these majorizers to converge more rapidly than
when using the less-structured diagonal majorizer $\M_\text{SQS}$.  Thanks to
the FFT, the circulant majorizer $\M_\text{Circ}$ is only marginally more
computationally expensive than $\M_\text{Diag}$.  On the other hand,
$\M_\text{Down}$ uses an approximation to the geometry of the original CT
system matrix that makes it less computationally expensive than $\A\tr\A$.
This allows $\M_\text{Down}$ to capture more of $\A\tr\A$'s structure than
$\M_\text{Circ}$, leading to even faster convergence.  While one could design a
circulant majorizer for $\A\tr\A$ using its point spread function and power
iteration, it is less clear how one would design a majorizer like
$\M_\text{Down}$ without an algorithm like the one proposed in this paper.

The ADMM-based reconstruction algorithms in this section may not converge
quickly in absolute terms, especially compared to contemporary fast CT
reconstruction
algorithms~\cite{nien:15:fxr,kim:15:cos,mcgaffin:15:fgd,mcgaffin::adu}, but
this experiment does show the benefits from using more sophisticated
majorizers.  Each
of~\cite{nien:15:fxr,kim:15:cos,mcgaffin:15:fgd,mcgaffin::adu} use diagonal
SQS-like majorizers, and they may benefit from algorithmically designed
majorizers.

\section{Conclusions and future work}

In this paper, we proposed a new way to design matrix majorizers.  Our
algorithm uses a duality-based approach to avoid relying on memory- and
computation-intensive semidefinite programming (SPD) techniques.  We proposed a
simple steepest ascent algorithm to find a local maximum of the nonconcave dual
problem, and we showed how to manipulate suboptimal solutions to guarantee a
majorizer.  In two preliminary experiments, we demonstrated the usefulness of
the algorithmically designed majorizers.  By capturing more of the structure
of the majorized matrix, algorithmically designed majorizers with appropriately
chosen structures can yield significant acceleration.

The experiments in this paper are preliminary, and future work will present
comparisons with state-of-the-art image reconstruction algorithms.  The
majorizer structure in this paper can also be used to design so-called
block-separable surrogates~\cite{kim:15:dbs} that may useful in distributed
computing.

\section*{Acknowlegments}

The authors acknowledge helpful discussions with Dr.\ Sathish Ramani
about majorizer design.

\bibliographystyle{plain}
\bibliography{master}

\begin{thebibliography}{10}

\bibitem{erdogan:99:osa}
H.~{\Erdogan} and J.~A. Fessler.
\newblock Ordered subsets algorithms for transmission tomography.
\newblock {\em Phys. Med. Biol.}, 44(11):{2835--51}, November 1999.

\bibitem{boyd:04}
S.~Boyd and L.~Vandenberghe.
\newblock {\em Convex optimization}.
\newblock Cambridge, UK, 2004.

\bibitem{chan:88:aoc}
T.~F. Chan.
\newblock An optimal circulant preconditioner for {Toeplitz} systems.
\newblock {\em SIAM J. Sci. Stat. Comp.}, 9(4):{766--71}, July 1988.

\bibitem{chouzenoux:11:amm}
E.~Chouzenoux, J.~Idier, and {\Said}~Moussaoui.
\newblock A majorize-minimize strategy for subspace optimization applied to
  image restoration.
\newblock {\em IEEE Trans. Im. Proc.}, 20(6):{1517--28}, June 2011.

\bibitem{clinthorne:93:pmf}
N.~H. Clinthorne, T.~S. Pan, P.~C. Chiao, W.~L. Rogers, and J.~A. Stamos.
\newblock Preconditioning methods for improved convergence rates in iterative
  reconstructions.
\newblock {\em IEEE Trans. Med. Imag.}, 12(1):{78--83}, March 1993.

\bibitem{deleeuw:09:sqm}
J.~{de Leeuw} and K.~Lange.
\newblock Sharp quadratic majorization in one dimension.
\newblock {\em Comp. Stat. Data Anal.}, 53(7):{2471--84}, May 2009.

\bibitem{fessler:93:ocd}
J.~A. Fessler, N.~H. Clinthorne, and W.~L. Rogers.
\newblock On complete data spaces for {PET} reconstruction algorithms.
\newblock {\em IEEE Trans. Nuc. Sci.}, 40(4):{1055--61}, August 1993.

\bibitem{fletcher:85:sdm}
R.~Fletcher.
\newblock Semi-definite matrix constraints in optimization.
\newblock {\em SIAM J. Cont. Opt.}, 23(4):{493--}, July 1985.

\bibitem{kim:15:dbs}
D.~Kim and J.~A. Fessler.
\newblock Distributed block-separable ordered subsets for helical {X-ray} {CT}
  image reconstruction.
\newblock In {\em Proc. Intl. Mtg. on Fully 3D Image Recon. in Rad. and Nuc.
  Med}, pages {138--41}, 2015.

\bibitem{kim:15:ogm}
D.~Kim and J.~A. Fessler.
\newblock Optimized gradient methods for smooth convex minimization.
\newblock In {\em Intl. Symp. Math. Prog.}, 2015.
\newblock Submitted.

\bibitem{kim:13:aos}
D.~Kim, D.~Pal, J-B. Thibault, and J.~A. Fessler.
\newblock Accelerating ordered subsets image reconstruction for {X-ray} {CT}
  using spatially non-uniform optimization transfer.
\newblock {\em IEEE Trans. Med. Imag.}, 32(11):{1965--78}, November 2013.

\bibitem{kim:15:cos}
D.~Kim, S.~Ramani, and J.~A. Fessler.
\newblock Combining ordered subsets and momentum for accelerated {X-ray} {CT}
  image reconstruction.
\newblock {\em IEEE Trans. Med. Imag.}, 34(1):{167--78}, January 2015.

\bibitem{mcgaffin::adu}
M.~G. McGaffin and J.~A. Fessler.
\newblock Alternating dual updates algorithm for {X}-ray {CT} reconstruction on
  the {GPU}.
\newblock {\em IEEE Trans. Comp. Imag.}, 2015.
\newblock To appear.

\bibitem{mcgaffin:15:fgd}
M.~G. McGaffin and J.~A. Fessler.
\newblock Fast {GPU-driven} model-based {X-ray} {CT} image reconstruction via
  alternating dual updates.
\newblock In {\em Proc. Intl. Mtg. on Fully 3D Image Recon. in Rad. and Nuc.
  Med}, pages {312--5}, 2015.

\bibitem{muckley:15:fpm}
M.~J. Muckley, D.~C. Noll, and J.~A. Fessler.
\newblock Fast parallel {MR} image reconstruction via {B1-based,} adaptive
  restart, iterative soft thresholding algorithms {(BARISTA)}.
\newblock {\em IEEE Trans. Med. Imag.}, 34(2):{578--88}, February 2015.

\bibitem{nesterov:83:amf}
Y.~Nesterov.
\newblock A method for unconstrained convex minimization problem with the rate
  of convergence {$O(1/k^2)$}.
\newblock {\em Dokl. Akad. Nauk. USSR}, 269(3):{543--7}, 1983.

\bibitem{nesterov:05:smo}
Y.~Nesterov.
\newblock Smooth minimization of non-smooth functions.
\newblock {\em Mathematical Programming}, 103(1):{127--52}, May 2005.

\bibitem{nien:15:fxr}
H.~Nien and J.~A. Fessler.
\newblock Fast {X-ray} {CT} image reconstruction using a linearized augmented
  {Lagrangian} method with ordered subsets.
\newblock {\em IEEE Trans. Med. Imag.}, 34(2):{388--99}, February 2015.

\bibitem{oettli:98:cff}
Werner Oettli and Dirk Schl\"ager.
\newblock Conjugate functions for convex and nonconvex duality.
\newblock {\em Journal of Global Optimization}, 13(4):337--347, 1998.

\bibitem{ramani:12:asb}
S.~Ramani and J.~A. Fessler.
\newblock A splitting-based iterative algorithm for accelerated statistical
  {X-ray} {CT} reconstruction.
\newblock {\em IEEE Trans. Med. Imag.}, 31(3):{677--88}, March 2012.

\bibitem{ramani:13:ans}
S.~Ramani and J.~A. Fessler.
\newblock Accelerated {nonCartesian} {SENSE} reconstruction using a
  majorize-minimize algorithm combining variable-splitting.
\newblock In {\em Proc. IEEE Intl. Symp. Biomed. Imag.}, pages {704--7}, 2013.

\bibitem{segars:08:rcs}
W.~P. Segars, M.~Mahesh, T.~J. Beck, E.~C. Frey, and B.~M.~W. Tsui.
\newblock Realistic {CT} simulation using the {4D} {XCAT} phantom.
\newblock {\em Med. Phys.}, 35(8):{3800--8}, August 2008.

\bibitem{thibault:07:atd}
J-B. Thibault, K.~Sauer, C.~Bouman, and J.~Hsieh.
\newblock A three-dimensional statistical approach to improved image quality
  for multi-slice helical {CT}.
\newblock {\em Med. Phys.}, 34(11):{4526--44}, November 2007.

\end{thebibliography}

\appendices

\renewcommand\x{\ensuremath{{\mathbf{z}} }}

\section{Strong duality}
\label{sec,strong,duality}

Let $\H \in \Complex^{N \times N} \succ \zeros$ be a given positive
semidefinite matrix, $\K \in \Complex^{K \times N}$, $K \ge N$ have full column
rank and define $\Omega$ to be
\begin{align}
    \Omega
    &=
    \GC{\d : \K\th\D\K \succeq \H}.
\end{align}
Consider the function
\begin{align}
    \mathsf{S}\GP{\d, \x} &= \frac{1}{2} \GN{\d}_\W^2 - \d\tr\GM{\K\x}^2 + \x\th\H\x.
\end{align}
The primal function is
\begin{align}
    \mathsf{J}\GP{\d} &= \sup_{\x \in \Complex^N} \mathsf{S}\GP{\d, \x} \\
                      &= \frac{1}{2} \GN{\d}_\W^2 + \iota_\Omega\GP{\d}.
\end{align}
and the dual function is
\begin{align}
    \mathsf{L}\GP{\x} &= \inf_{\d \in \Reals^K} \mathsf{S}\GP{\d, \x} \\
                      &= - \frac{1}{2}\GN{\GM{\K\x}^2}_{\W^{-1}}^2 + \x\th\H\x.
\end{align}
In this section we show the minimum of the primal function is equal to the
maximum of the dual function:
\begin{align}
    p = \min_{\d \in \Reals^K} \mathsf{J}\GP{\d} = \sup_{\x \in \Complex^N} \mathsf{L}\GP{\x} = d.
\end{align}

Because $\mathsf{J}$ is a strongly convex function and $\Omega$ is a convex
set, there exists a unique minimizer of $\mathsf{J}$ over $\Omega$.  Let
$\widehat \d$ be this minimizer:
\begin{align}
    \widehat \d = \argmin_{\d \in \Omega} \mathsf{J}\GP{\d}.
\end{align}
Because $\H \succ \zeros$ the unconstrained minimizer of
$\frac{1}{2}\GN{\d}_\W^2$, $\d_\text{unconstrained} = \zeros$, does not lie in
$\Omega$.  Therefore, $\widehat \d$ is on the boundary of $\Omega$ and $-\grad
\mathsf{J}\GP{\widehat \d} = \W \widehat \d$ is normal to $\Omega$.

We can characterize the feasible set $\Omega$ as an intersection of
half-spaces:
\begin{align}
    \Omega
    &=
    \bigcap_{\x} \GC{\d: \x\th\GP{\K\th\D\K - \H}\x \ge 0} \\
    &=
    \bigcap_\x \GC{\d: \d \tr \GM{\K\x}^2 \ge \x\th\H\x}.
\end{align}
Because $-\grad\mathsf{J}\GP{\widehat \d} = \W \widehat \d$ is normal to $\Omega$
and on the boundary of $\Omega$, there exists an $\widehat \x$ for which
one of the above inequalities holds with equality.  That is,
\begin{align}
    \elabel{proof,a}
    -\grad \mathsf{J}\GP{\widehat \d} = \W \widehat \d = \alpha\GM{\K\widehat\x}^2
\end{align}
and
\begin{align}
    \elabel{proof,b}
    \widehat \x \th \H \widehat \x
        = \widehat \d \tr \GM{\K\widehat\x}^2
        = \alpha \GP{\GM{\K\widehat\x}^2}\tr\W^{-1}\GM{\K\widehat\x}^2
        = \alpha \GN{\GM{\K\widehat \x}^2}_{\W^{-1}}^2.
\end{align}
We use~\eref{proof,a} to find the minimum of the primal function:
\begin{align}
    \mathsf{J}\GP{\widehat \d}
    &= \frac{1}{2}\GN{\widehat \d}_\W^2 \\
    &= \frac{1}{2} \GN{\alpha \W^{-1} \GM{\K\widehat \x}^2}_\W^2 \\
    &= \frac{\alpha^2}{2} \GN{\GM{\K\widehat \x}^2}_{\W^{-1}}^2 \\
    &= p.
\end{align}

It is widely known that $p \ge d$.  Therefore, to show $p = d$ it suffices to
show that $\mathsf{L}\GP{\x} = p$ for some $\x$.  We try $\mathsf{L}\GP{\beta
\widehat \x}$, with $\beta^2 = \alpha$:
\begin{align}
    \mathsf{L}\GP{\beta \widehat\x}
    &=
    - \frac{1}{2}\GN{\GM{\beta \K \widehat \x}^2}_{\W^{-1}}^2 + \beta^2 \widehat \x \tr \H \widehat \x \\
    &=
    - \frac{\beta^4}{2}\GN{\GM{\K \widehat \x}^2}_{\W^{-1}}^2 + \beta^2 \widehat \x \tr \H \widehat \x
    \intertext{via~\eref{proof,b},}
    &=
    - \frac{\beta^4}{2}\GN{\GM{\K \widehat \x}^2}_{\W^{-1}}^2 + \beta^2 \alpha \GN{\GM{\K\widehat\x}^2}_{\W^{-1}}^2 \\
    &=
    \frac{\alpha^2}{2} \GN{\GM{\K\widehat\x}^2}_{\W^{-1}}^2 \\
    &= p,
\end{align}
completing the proof.

\section{Equivalence of majorizers from primal and dual problems}
\label{sec,recover}

In this paper, instead of solving the primal problem
\begin{align}
    \widehat \D &= \argmin_{\d \in \Reals^K} \frac{1}{2}\GN{\d}_\W^2 + \iota_\Omega\GP{\d} \\
                &= \argmin_{\d \in \Reals^K} \sup_{\x \in \Complex^N} \frac{1}{2}\GN{\d}_\W^2 + \x\th\H\x - \d\tr\GM{\K\x}^2 \\
                &= \argmin_{\d \in \Reals^K} \sup_{\x \in \Complex^N} \mathsf{S}\GP{\d, \x},
\end{align}
we reverse the order of the minimization and maximization and solve the dual
problem
\begin{align}
    \elabel{proof,dual,solution}
    \widehat \x &= \argmax_{\x \in \Complex^N} \inf_{\d \in \Reals^K} \mathsf{S}\GP{\d, \x}.
\end{align}
In this section, we prove that the primal solution
\begin{align}
    \widehat \D_p &= \argmin_{\d \in \Reals^K} \sup_{\x \in \Complex^N} \mathsf{S}\GP{\d, \x}
\end{align}
and the solution induced by solving the dual problem
\begin{align}
    \widehat \D_d &= \argmin_{\d \in \Reals^K} \mathsf{S}\GP{\d, \widehat \x},
\end{align}
where $\widehat \x$ solves the dual problem~\eref{proof,dual,solution}, are
equal.

Let $d$ be the maximum value attained by the dual function at $\widehat \x$
and $p$ be the minimum value attained by the primal function at $\widehat \d$:
\begin{align}
    d &= \sup_{\x \in \Complex^N} \inf_{\d \in \Reals^K} \mathsf{S}\GP{\d, \x}, \\
    p &= \inf_{\d \in \Reals^K} \sup_{\x \in \Complex^N} \mathsf{S}\GP{\d, \x}.
\end{align}
We proceed by contradiction.  Assume that $\widehat \D_p \ne \widehat \D_d$.
Since $\mathsf{S}\GP{\d, \x}$ is a strongly convex function of $\d$
\begin{align}
    d &= \sup_{\x \in \Complex^N} \inf_{\d \in \Reals^K} \mathsf{S}\GP{\d, \x} \\
      &= \mathsf{S} \GP{\d_d, \widehat \x} \\
      &< \mathsf{S} \GP{\d_p, \widehat \x} \\
      &\le \mathsf{S} \sup_{\x \in \Complex^N} \GP{\d_p, \x} \\
      &= p.
\end{align}
That is, $p \ne d$.  This contradicts the strong duality result in
Section~\ref{sec,strong,duality}, and we conclude that $\widehat \D = \D_p =
\D_d$.  Now we can write the primal solution $\widehat \D$ in terms of dual
solution $\widehat \x$:
\begin{align}
    \widehat \D
    &=
    \argmin_{\d \in \Reals^K} \mathsf{S}\GP{\d, \widehat \x} \\
    &=
    \argmin_{\d \in \Reals^K} \frac{1}{2}\GN{\d}_\W^2 - \d\tr\GM{\K\d}^2 + \x\th\H\x \\
    &=
    \W^{-1}\GM{\K\d}^2.
\end{align}

\section{Scaling for majorization}
\label{sec,scaling}

Let $\widetilde \x$ be a local maximum of the nonconcave dual function
$\mathsf{L}\GP{\x}$ found by an iterative gradient-based method.  Because
$\widetilde \x$ is an attractor of the maximization procedure and $\mathsf{L}$
is smooth, $\mathsf{L}$ is concave at $\widetilde \x$.  That is, the Hessian
of $\mathsf{L}$ at $\widetilde \x$ is nonpositive definite:
\begin{align}
    \hess \mathsf{L}\GP{\widetilde \x}
    &=
    - 6 \K\th \W^{-1} \diag_{i}\GC{\GB{\GM{\K\widetilde\x}^2}_i} \K + 2 \H \preceq \zeros.
\end{align}
Rearranging,
\begin{align}
    3 \K\th \W^{-1} \diag_{i}\GC{\GB{\GM{\K\widetilde\x}^2}_i} \K &\succeq \H.
\end{align}
Even though we do not find the global maximum of $\mathsf{L}$ using the
steepest ascent procedure in Section~\ref{sec,solve}, simply scaling the
majorizer produced by a local maximum by a factor of 3 produces a majorizer
for $\H$.

\end{document}